\documentclass[11pt]{article}

\usepackage{a4wide}
\usepackage{xspace}
\usepackage{multicol}
\usepackage{amsmath}
\usepackage{graphicx}
\usepackage{epsfig}
\usepackage{hyperref} %linkable table of contents, eqns, refs...
\hypersetup{linktocpage} %link only the number
\hypersetup{
    colorlinks,
    citecolor=blue,
    filecolor=blue,
    linkcolor=blue,
    urlcolor=blue}
\usepackage{amssymb}

\newcommand{\ycut}{y_\text{cut}}
\newcommand{\zcut}{z_\text{cut}}
\newcommand{\rhofat}{\rho_\text{fat}}

\newcommand{\Rfact}{R_\text{fact}}
\newcommand{\Rprune}{R_\text{prune}}
\newcommand{\Rsub}{R_\text{sub}}

\newcommand{\nfilt}{n_\text{filt}}

\newcommand{\TeV}{\,\mathrm{TeV}}

\newcommand{\as}{\alpha_s}
\newcommand{\order}[1]{{\cal O}\left(#1\right)}
\newcommand{\sanepruning}{\text{Y-pruning}}

\newcommand{\saneprune}{\text{(Y-prune)}}

\newcommand{\sane}{\text{Y}}
\newcommand{\anomalouspruning}{\text{I-pruning}}

\newcommand{\anomprune}{\text{(I-prune)}}

\newcommand{\anomalous}{\text{I}}

\newcommand{\beq}{\begin{equation}}
\newcommand{\eeq}{\end{equation}}
\newcommand{\bea}{\begin{eqnarray}}
\newcommand{\eea}{\end{eqnarray}}
\newcommand{\bdm}{\begin{displaymath}}
\newcommand{\edm}{\end{displaymath}}
\def\as{\alpha_s}

\def\d{\partial}

\def \d{{\rm d} }
\def \d0 {D\O \;}

\title{\textbf{QCD calculations for jet substructure~\footnote{Talk
      presented at various conferences including: ESI Program on Jets
      and QFT, Beyond the LHC Nordita Workshop, Boost 2013, QCD@LHC
      2013, LC 2013 and Radcor 2013.}}} 
\author{
  Mrinal Dasgupta,\!$^{1}$ Simone Marzani$^{2}$ and Gavin P.~Salam$^{3,4}$ \\
\\
{\sl  \small $^1$Consortium for Fundamental Physics, School of Physics \& Astronomy,} \\ {\sl \small University of Manchester, Manchester M13 9PL, United Kingdom }\\[2pt]
{\sl  \small $^2$Institute for Particle Physics Phenomenology,}\\ {\sl \small 
Durham University, Durham DH1 3LE, United Kingdom}\\[2pt]
{\sl  \small $^3$CERN, PH-TH, CH-1211 Geneva 23, Switzerland}\\[2pt]
{\sl  \small $^4$LPTHE; CNRS UMR 7589; UPMC Univ. Paris 6; Paris 75252, France}}
\date{}

\begin{document}

\maketitle

%\vspace{-10.0cm}
% \begin{flushright}
%   DCPT/13/88 \\ IPPP/13/44 \\ MAN/HEP/2013/11
%%
% \end{flushright}
%\vspace{8cm}

\begin{abstract}
We present results on novel analytic calculations to describe invariant mass distributions of QCD jets with three substructure algorithms: trimming, pruning and the mass-drop taggers. These results not only lead to considerable insight into the behaviour of these tools, but also show how they can be improved. As an example, we discuss the remarkable properties of the modified mass-drop tagger.
\end{abstract}

%
%
%
% 
% 

%\clearpage

%\tableofcontents
%\newpage

%
%%%%%%%%%%%%%%%%%%%%%%%%%%%%%%%%%%%%%%%%%%%%%%%%%%%%%%%%%%%%%%%%

\section{The boosted regime and substructure tools}\label{sec:intro}

The main aims of the research programme carried out at the the Large Hadron Collider (LHC) at CERN are to understand the mechanism of electroweak symmetry breaking and to explore the TeV scale for signs of new physics beyond the Standard Model of particle physics.
In order to achieve this, protons are collided at energies far above the electroweak scale, opening up the possibility of producing 
 electroweak-scale particles with a large boost.
In these situations, their hadronic decay products are collimated into a single jet.
Consequently a vibrant research field has emerged in recent years,
investigating how best to identify the characteristic substructure that
appears inside ``signal" jets in order to differentiate them from background (QCD) jets (for a review of the field see Refs~\cite{Boost2010, Boost2011, Boost2012, PlehnSpannowsky}).
Many ``grooming" and ``tagging" algorithms have been developed, successfully tested and are already being used experimental analyses
(in particular see Refs~\cite{ATLAS:2012am,Aad:2012meb,Aad:2013gja,CMS-substructure-studies} for
studies on QCD jets).

Until very recently, nearly all the theoretical studies of substructure tools 
have been done using Monte Carlo parton showers. While these are powerful general purpose tools, their essentially numerical nature offers little insight into the results produced or their detailed and precise dependence on tagger parameters and the parameters of jet finding. Such a detailed level of understanding, which can be achieved for example via  analytical formulae, is in fact crucial in order for substructure studies to realise their full potential. However it has been {\it{far from obvious}} that, given their inherent complexity, substructure taggers can be understood to any extent analytically.

In two recent papers~\cite{DasFregMarSal,DasFregMarPow} we have developed the 
first comprehensive theoretical understanding of three commonly used substructure tools: trimming~\cite{Krohn:2009th}, pruning~\cite{Ellis:2009me,Ellis:2009su} and the mass drop tagger~\cite{Butterworth:2008iy}. In these proceedings we review the main results of those papers, focussing on the perturbative 
properties of jet mass distributions of QCD jets with the application of substructure algorithms, and compare the results to the plain jet mass distribution.  

\section{The perturbative structure of jet mass distributions}\label{sec:qcdjets}

Jet mass distributions are affected by logarithmic corrections in the ratio of jet invariant mass ($m$) over its transverse momentum ($p_t$). When this ratio becomes small, as happens for highly boosted configurations, these logarithms are large and fixed-order perturbation theory is not a reliable way to organise the calculation . One then needs to resum these large corrections to all orders in perturbation theory. Resummed results can be then matched to fixed-order ones, typically obtained at next-to-leading order (NLO), in order to obtain a reliable estimate of jet masses over a wide range of $m/ p_t$.

\subsection{Plain jet mass}\label{sec:plain-mass}
Resummed calculations are usually discussed in terms of the cumulative distributions, i.e.\ the integral of the jet mass distribution up to a fixed value:
\begin{equation}\label{cum-def}
\Sigma(\rho) = \frac{1}{\sigma} \int^\rho \frac{d \sigma}{d \rho'} \, d \rho', \quad \rho= \frac{m^2}{p_t^2 R^2},
\end{equation}
where $R$ is the jet radius. In our discussion, we will work in the small jet radius limit $R \ll 1$. This considerably simplifies our expressions because 
we only have to consider the radiation from the parton that initiated the jet: large-angle radiation from other final-state partons and from the initial-state partons result in contributions which are power suppressed in $R$. For brevity, we also limit ourselves to the case of quark-initiated jets.

To next-to-leading logarithmic (NLL) accuracy, i.e.\ control of terms $\as^n
L^{n+1}$ and $\as^n L^n$ in $\ln \Sigma(\rho)$, where $L \equiv \ln
\frac1{\rho}$, the cumulative distribution can be computed using an independent-emission approximation, ignoring
subsequent splittings of those emissions, other than in the treatment
of the running coupling (see for instance~\cite{Catani:1992ua,Dokshitzer:1998kz}) and of non-global contributions~\cite{Dasgupta:2001sh}. The NLL result, in the small-$R$ limit, can be written as
\begin{equation}
  \label{eq:Sigma-plain-jet-mass}
  \Sigma(\rho) = e^{-D(\rho)} \cdot \frac{e^{-\gamma_E
     D'(\rho)}}{\Gamma(1 +D'(\rho))} \cdot \mathcal{N}(\rho)\,.
\end{equation}
The first factor, which is double logarithmic, accounts for the
Sudakov suppression of emissions that would induce a (squared,
normalised) jet mass greater than $\rho$. In a fixed coupling approximation the resummed exponent reduces to
  \begin{equation} \label{D-def}
    D(\rho) \simeq \frac{\as C_F}{\pi} \left[ \frac12 \ln^2 \frac1\rho \;-\;
     \frac34 \ln \frac1\rho + \order{1}\right],
     \end{equation}
     The second factor in Eq.~(\ref{eq:Sigma-plain-jet-mass}), accounts for the fact that the
effects of multiple emissions add together to give the jet's overall
mass.
The third factor, also single logarithmic, accounts for modifications
of the radiation pattern in the jet (non-global
logarithms~\cite{Dasgupta:2001sh}) and boundaries of the jet
(clustering logarithms~\cite{Appleby:2002ke,DasBan05,Delenda:2006nf,Kelley:2012kj}) induced
by soft radiation near the jet's edge.

Non-global logarithms are the main obstacle to a full resummation of the standard jet mass beyond NLL accuracy (for
work towards higher accuracy, see Refs.~\cite{Chien:2012ur,Jouttenus:2013hs})
and why even the NLL calculations have to neglect $1/N_C^2$ suppressed terms, as done in
Ref.~\cite{Banfi:2010pa, Dasgupta:2012hg}~\footnote{Resummation of non-global logarithms with full $N_C$ dependence has been recently achieved in Ref.~\cite{Hatta:2013iba}.}.

\subsection{Trimmed mass distribution}\label{sec:trimming}

Trimming~\cite{Krohn:2009th} takes all the particles in a jet of radius $R$ and reclusters
them into subjets with a jet definition with radius $\Rsub < R$.
All resulting subjets that satisfy the condition $p_t^\text{(subjet)}
> \zcut p_t^\text{(jet)}$ are kept and merged to form the trimmed
jet. The other subjets are discarded.

We can get  an idea of the trimmed jet mass behaviour by considering configurations in which the jet is made of a hard quark and a bunch of soft gluons. It is then clear that the algorithm will cut away soft radiation if emitted at angles larger than $\Rsub$, while arbitrarily soft gluons radiated at angles smaller than $\Rsub$ will contribute to the trimmed jet mass.

The full leading logarithmic (LL) calculation of the trimmed jet mass produces:
\begin{multline}
  \label{eq:trimming-LL-small-mass}
  \Sigma^\text{(trim)} (\rho)
  =
  \exp\Bigg[ 
    - D(\max(\zcut,\rho)) 
    - S(\zcut,\rho) \Theta(\zcut - \rho)
    \\ \left.
    - \Theta(\zcut r^2 - \rho) \int_\rho^{\zcut r^2} \frac{d\rho'}{\rho'}
    \int^{\zcut}_{\rho'/r^2} \frac{dz}{z} \frac{C_F}{\pi}\, \as(\rho' z p_t^2 R^2)
  \right]\,.
\end{multline}
where $r=\frac{\Rsub}{R}$ and we have neglected finite $\zcut$ corrections.
The resummed exponent $D$  is the same as in Eq.~(\ref{D-def}), while the function $S$ is single-logarithmic and in a fixed-coupling approximation is given by
\begin{equation}\label{S-def}
   S(a,b)\simeq \frac{\as C_F}{\pi} \left[ \ln \frac{1}{\zcut} - \frac34 +
      \order{\zcut} \right] \ln \frac{a}{b},
      \end{equation}
 
We can now discuss differences and similarities of the trimmed mass distribution in Eq.~(\ref{eq:trimming-LL-small-mass}) to the plain jet mass Eq.~(\ref{eq:Sigma-plain-jet-mass}). The main similarity from the point of view of resummed calculations is that in both cases the analysis of the one loop case essentially captures the LL behaviour to all orders (this is not the case for pruning or mass drop). 
However, the actual form of the one-gluon exponentiation in the case of trimming has a non-trivial dependence on the jet's kinematics.
We can identify three distinct kinematic regions: for $\rho>\zcut$ trimming is not active and the result is the same as plain jet mass. For  $r^2 \zcut < \rho<\zcut$, the parameter $\zcut$ provides a lower limit for the emissions' energy, resulting into a single-logarithmic distributions. The last region $\rho< r^2 \zcut$ is again double logarithmic and it correspond to configurations in which soft gluons are emitted at angles smaller than $\Rsub$, as mentioned above.

Eq.~(\ref{eq:trimming-LL-small-mass}) does not capture full NLL accuracy  i.e.\ all
terms $\as^n L^{n}$ in $\ln \Sigma(\rho)$.
The missing terms include non-global logarithms, related clustering logarithms, and multiple-emission
effects on the observable. They should all be relatively straightforward to include, if desired, since they follow the structure of corresponding terms for 
the plain jet-mass distribution.

\begin{figure}
 \begin{center}
\includegraphics[width=0.4 \textwidth]{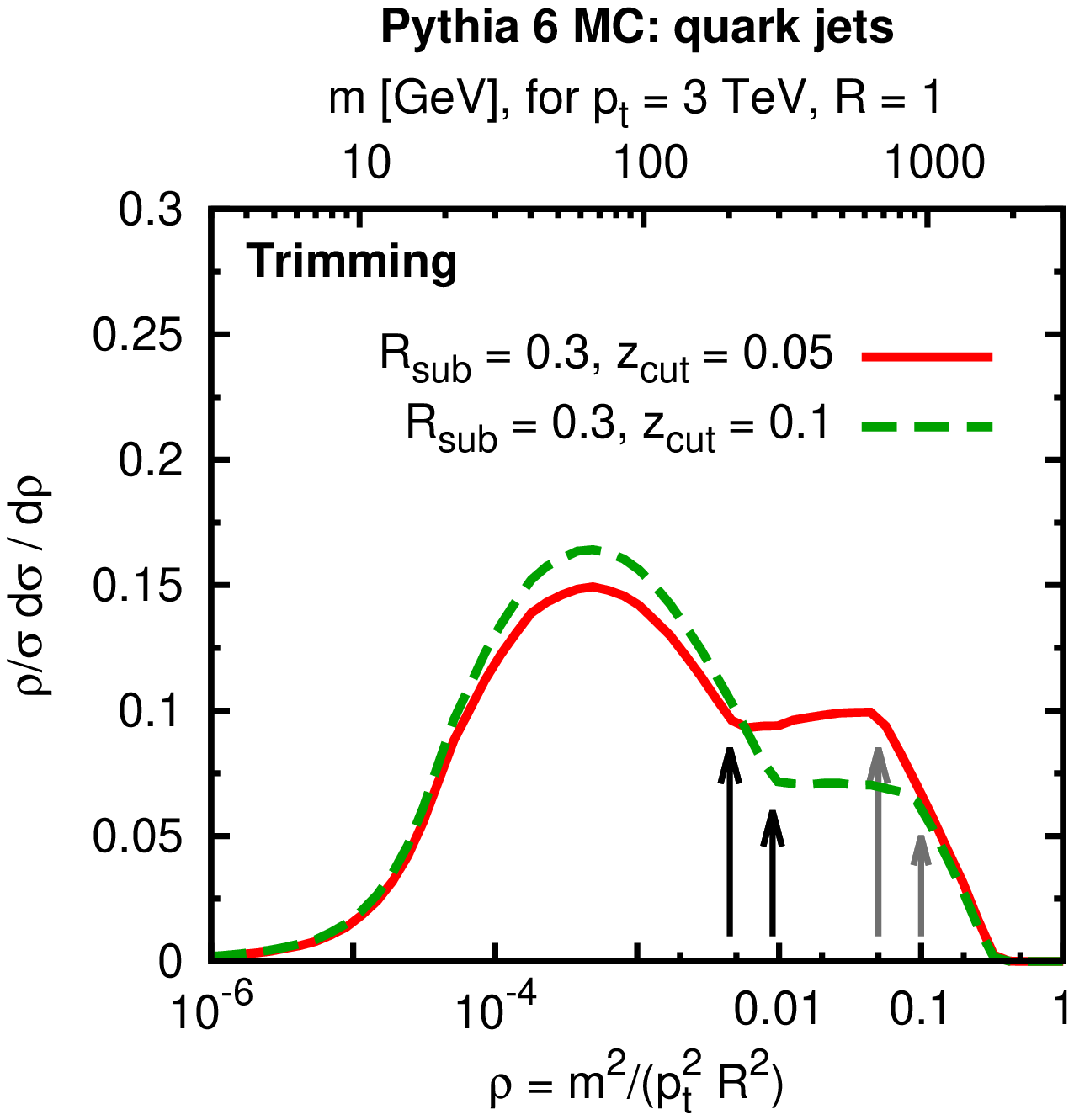}
\includegraphics[width=0.4 \textwidth]{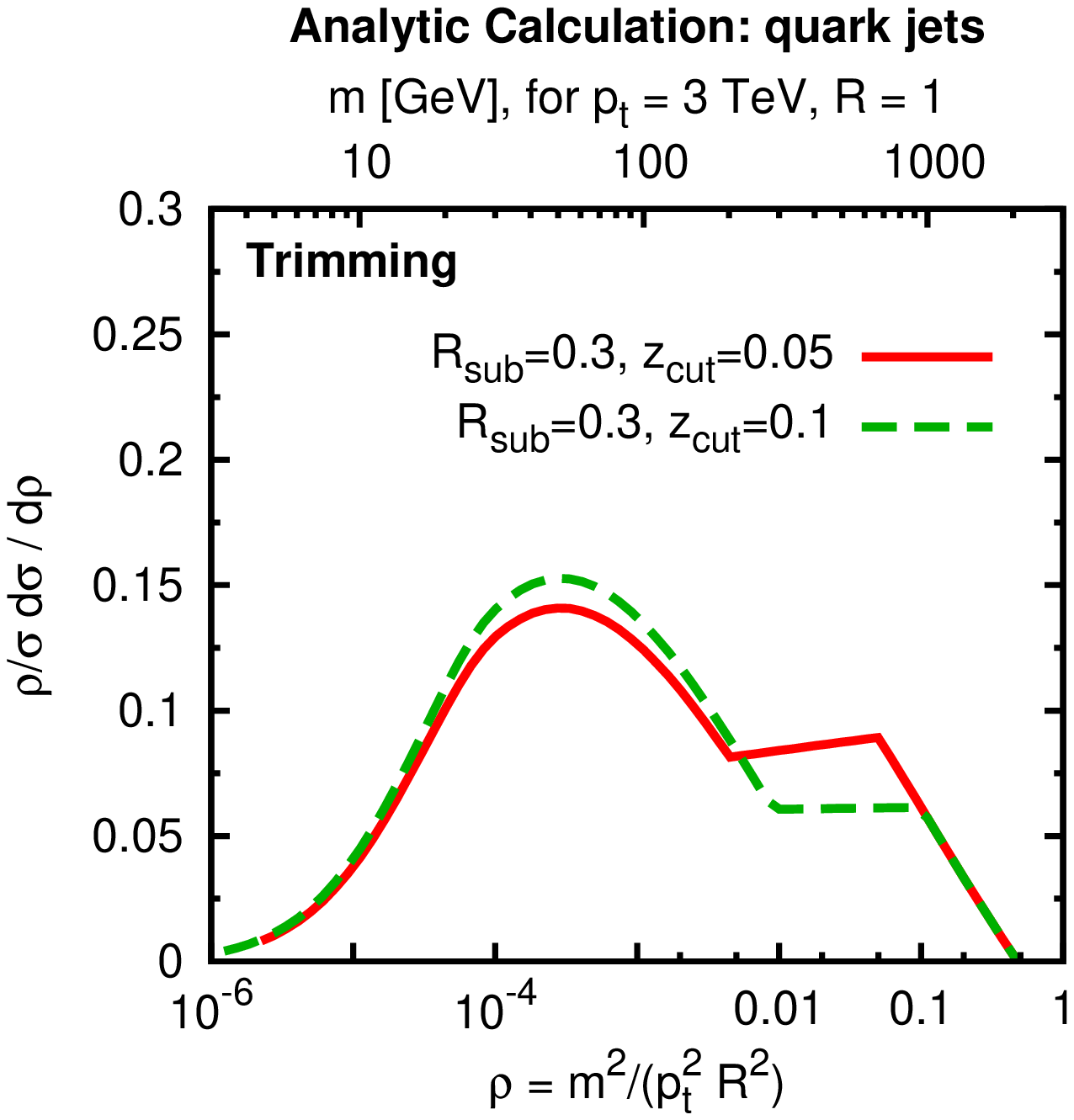}
\includegraphics[width=0.4 \textwidth]{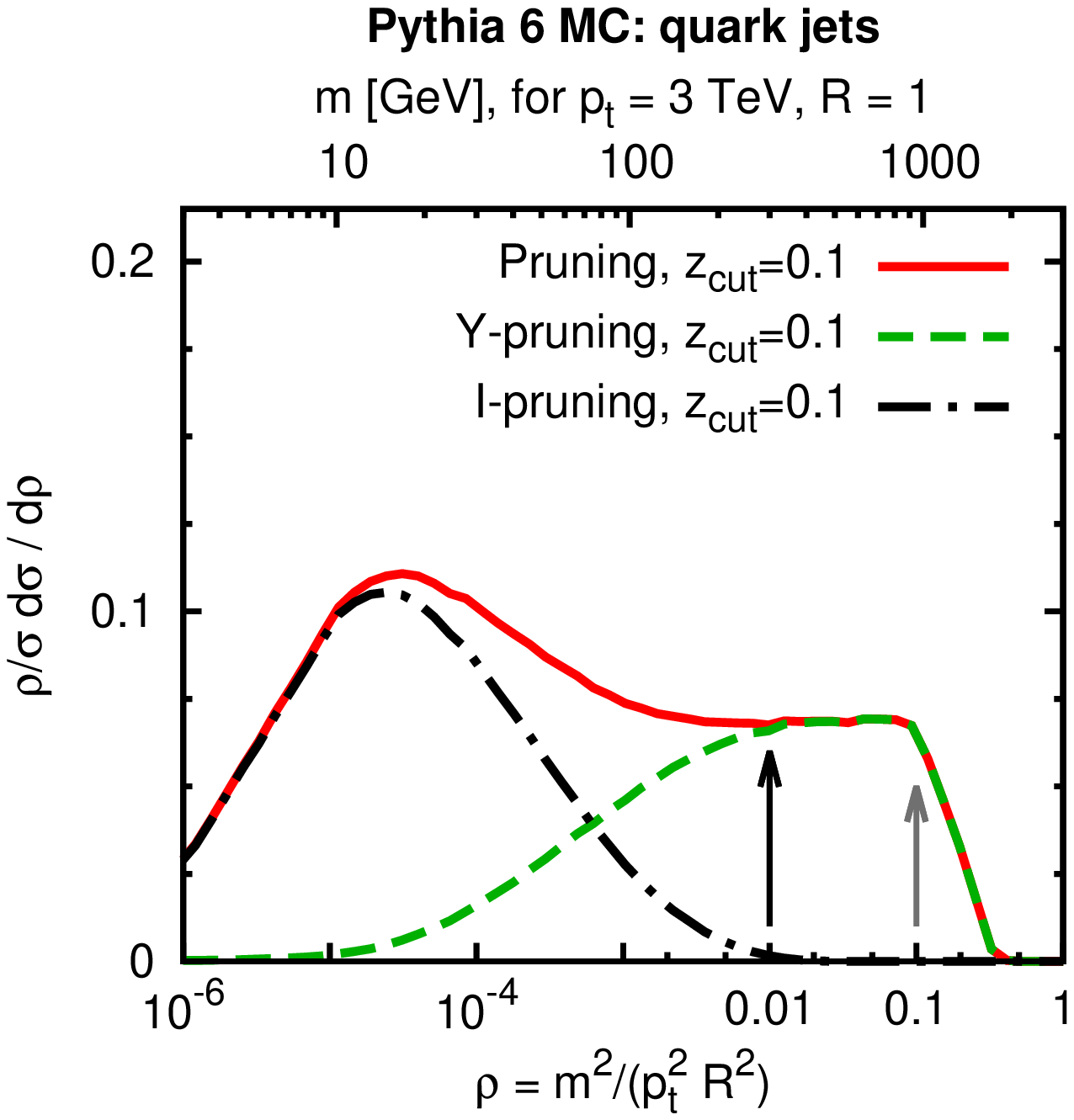}
\includegraphics[width=0.4 \textwidth]{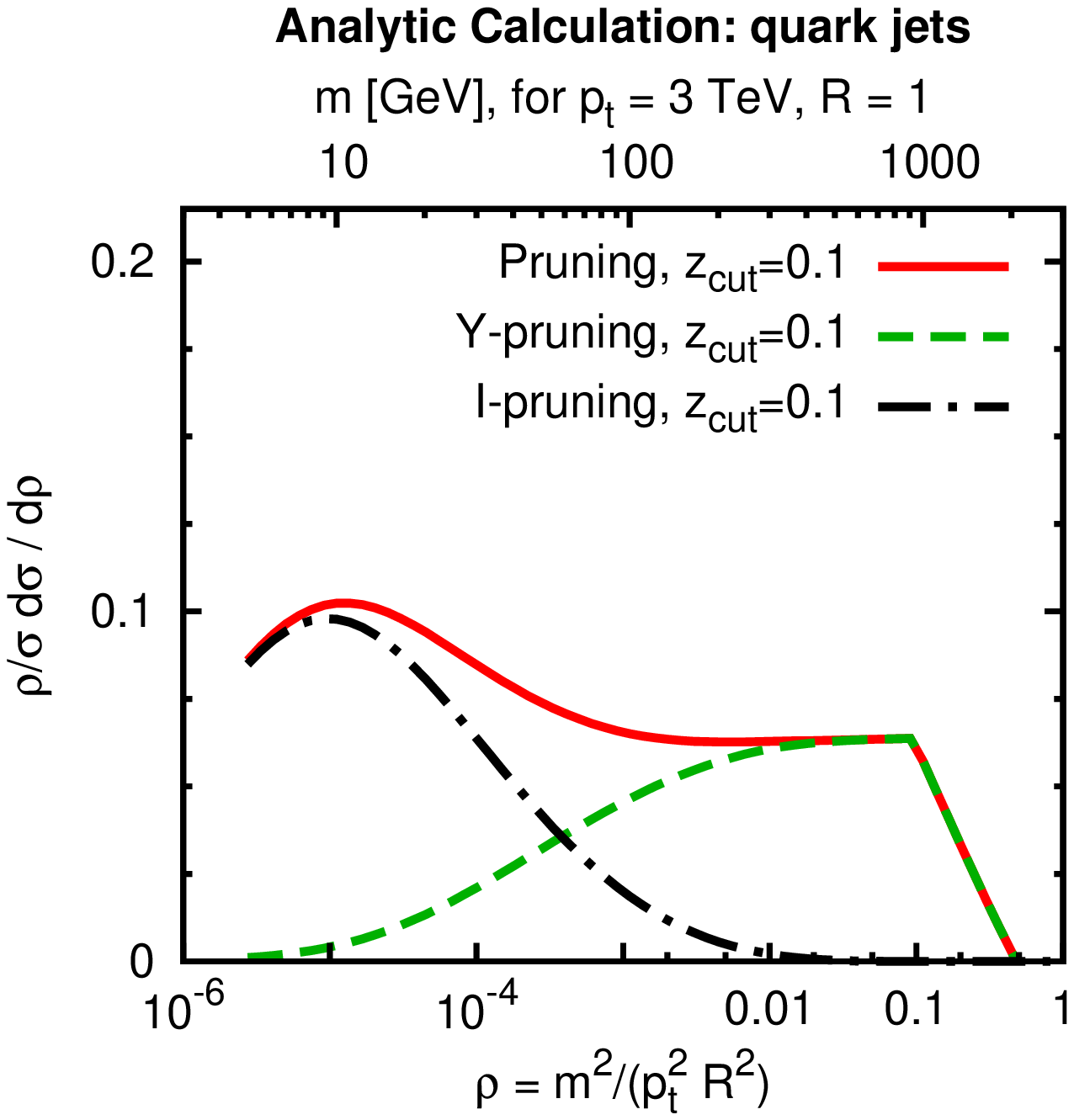}
\includegraphics[width=0.4 \textwidth]{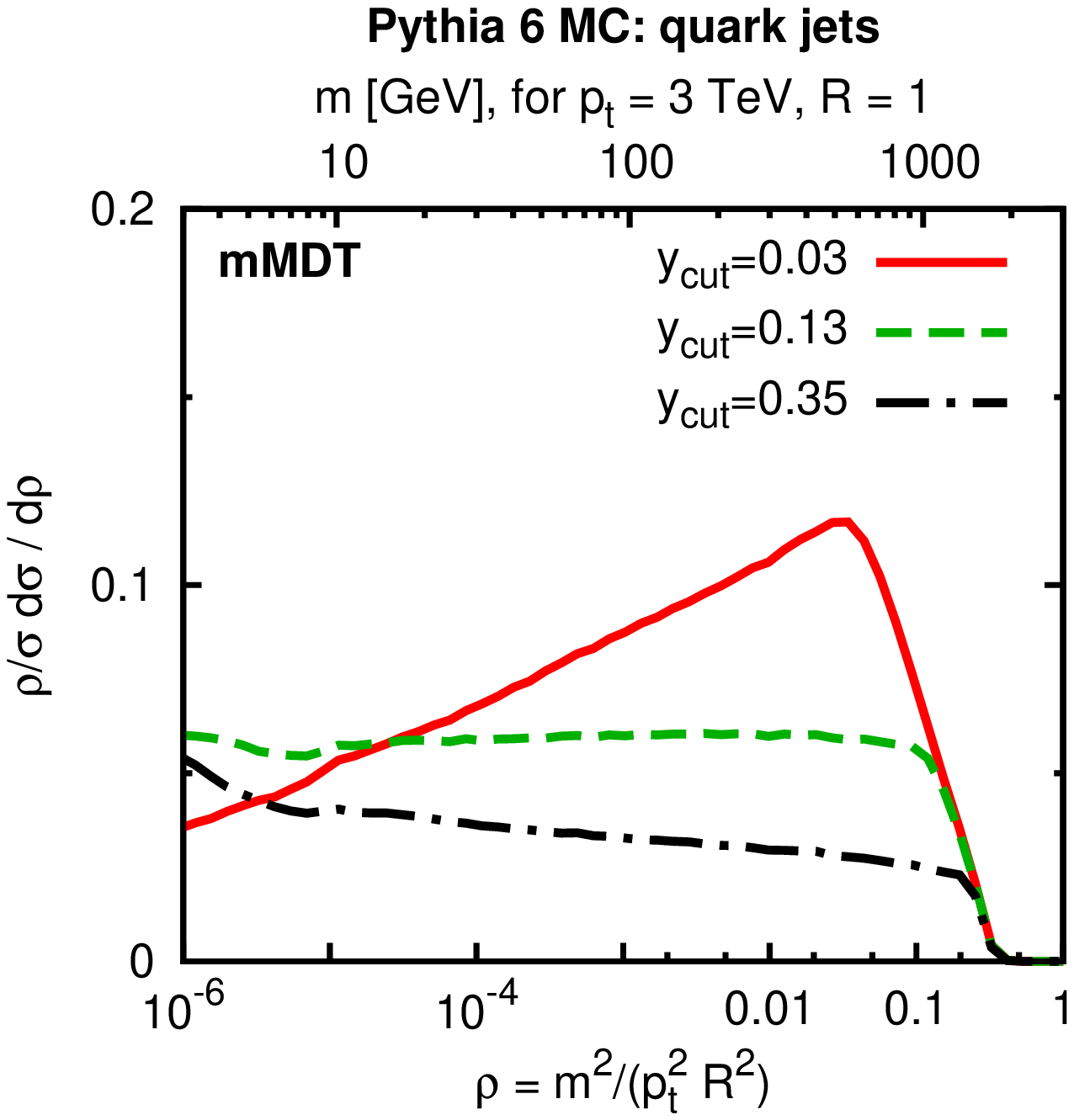}
\includegraphics[width=0.4 \textwidth]{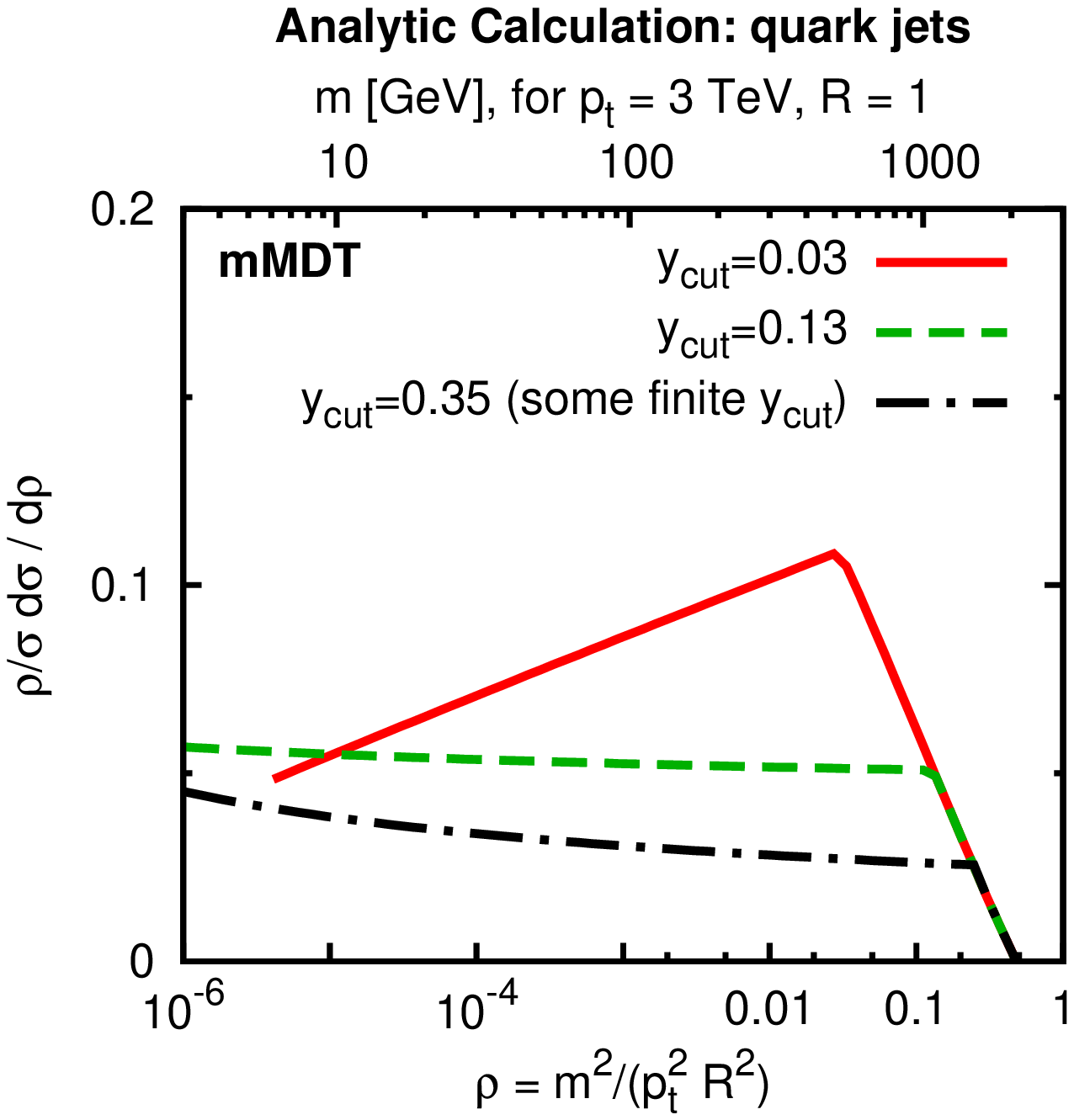}
\caption{Comparisons of the resummed calculations for the mass distributions (right-hand panels) to a standard parton shower (left-hand plots). 
The arrows indicate the analytic prediction for the position of the transition points.
     The results on the left-hand panels have been  obtained from Monte Carlo simulation
    with Pythia~6.425~\cite{Pythia6} in the DW tune~\cite{DW}
    (virtuality-ordered shower), with a minimum $p_t$ cut in the
    generation of $3\TeV$, for $14\TeV$ $pp$ collisions, at parton
    level, including initial and final-state showering, but without
    the underlying event (multiple interactions).  
    Clustering was performed with FastJet~\cite{Cacciari:2011ma}.
   }
\label{fig:analytic-vs-MC}
\end{center}
\end{figure}

In order to test that the approximations made in order to obtain the resummed result in Eq.~(\ref{eq:trimming-LL-small-mass}) capture the relevant physical effects, we compare our result to the one obtained with a Monte Carlo parton shower.
This comparison is shown in Fig.~\ref{fig:analytic-vs-MC}. Our calculations indeed reproduce the shape of the distribution in all three distinct regions, as well as the position of the transition points between these regions (indicated by vertical arrows), which confirms that we have analytically captured the 
essence of trimming.

\subsection{Pruned mass distribution}
Pruning~\cite{Ellis:2009me,Ellis:2009su} takes an initial jet, and
from its mass deduces a pruning 
radius $R_\text{prune} = \Rfact \cdot \frac{2m}{p_t}$, with $\Rfact$
of order 1 (here we adopt the widespread choice $\Rfact=0.5$, but our main conclusions do not depend on this 
choice).
It then reclusters the jet and for every clustering step, involving
objects $a$ and $b$, it checks whether $\Delta_{ab} > R_\text{prune}$
and $\min(p_{ta},p_{tb}) < \zcut p_{t,(a+b)}$, where $\zcut$ is a second
parameter of the tagger.
If so, then the softer of the $a$ and $b$ is discarded.
Otherwise $a$ and $b$ are recombined as usual.
Clustering then proceeds with the remaining objects, applying the
pruning check at each stage.

We can start our analysis by considering the $\order{\as}$ contribution.
At this perturbative order the jet is made of two partons, $a$ and $b$ with $m=m_{ab}$ and $\Delta_{ab} > R_\text{prune}=\frac{m_{ab}}{p_t}$. Thus, the two partons are kept only if they pass the energy condition, irrespectively of their angular distance. This has a remarkable consequence: the LO pruned mass distribution receives no contributions from the soft region and it has only a single logarithm, which is of pure collinear origin.

This behavior certainly appears desirable from the viewpoint of taming the background jet mass distribution as it rids us of double logarithms. Thus we may wonder if the above feature holds to all orders. However an analysis of the NLO 
contributions reveal that this is not the case and the pruned mass distribution receives contributions from soft emissions beyond LO and consequent double logarithmic enhancements appear.

In particular,  we consider NLO configurations (which involve three partons), 
where there is a soft parton ($p_3$) that dominates the total jet mass thus setting the pruning radius, but is soft enough that it fails the $\zcut$ threshold and therefore it does not contribute to the pruned mass; meanwhile there is another parton ($p_2$), within the pruning radius, that contributes to the pruned jet mass independently of how soft it is.
We call this ``$\anomalouspruning$'', because at the angular scale
$\Rprune$ (set by the soft parton $p_3$), the final pruned jet consists of a single hard prong.
On the other hand, we call ``$\sanepruning$'' those configurations that contributed to the leading order result  for which at an
angular scale $\Rprune$, the pruned jet always consisted of two prongs. 
  
The above analysis can be generalised to all orders and a resummed result for 
pruning and its $\sanepruning$ and $\anomalouspruning$ components can
be found. Here we report only simplified (double logarithmic) versions of the results, valid at fixed coupling, and we refer the reader to Ref.~\cite{DasFregMarSal} for more complete expressions. We have
\begin{equation}
\frac{d\sigma^{(\text{prune})}}{d\rho}=\frac{d\sigma^{\saneprune}}{d\rho}+\frac{d\sigma^{\anomprune}}{d\rho}\,,
\end{equation}
with
  \begin{eqnarray}
    \label{eq:prune-fixed-coupling} 
     \frac{\rho}{\sigma} \frac{d\sigma^{\saneprune}}{d\rho}   &\simeq& e^{-D(\rho)}\, \frac{\as C_F}{\pi} \left[\ln \frac{1}{\zcut}
      - \frac{3}{4} \right]\,,\nonumber\\
    \frac{\rho}{\sigma} \frac{d\sigma^{\anomprune}}{d\rho}  
   &\simeq&
     \left(\frac{\as C_F}{\pi}\right)^2 \int_\rho^1  
    \frac{d\rhofat}{\rhofat} \ln \rhofat e^{- \frac12 \frac{\as
        C_F}{\pi} \ln^2 \frac{1}{\rhofat}}
    \,
    \ln \frac{\rho}{\rhofat} e^{- \frac12 \frac{\as
        C_F}{\pi} \ln^2 \frac{\rhofat}{\rho}}\,.
\end{eqnarray}

Several comments can be made about the perturbative structure of the above results. First of all we note that the $\anomalouspruning$ distribution contains a convolution between two exponentials. The resulting distribution is double logarithmic, i.e. $\Sigma^\anomprune(\rho)$ contains $\as^n L^{2n}$ contributions and hence it is as singular as plain jet mass.
On the other hand, $\sanepruning$ is essentially a Sudakov suppression of the leading order result and, therefore, $\Sigma^\saneprune(\rho)$ is as singular as $\as^n L^{2n-1}$. Interestingly, when considering full pruning, i.e.\ the sum of the two components, a cancellation occurs in the $\zcut^2<\rho<\zcut$ region and one obtains a distribution which is only single-logarithmic.

As for trimming, to reach full NLL accuracy  would require the treatment
of several additional effects: non-global logarithms and related
clustering logarithms and multiple-emission effects on the observable.
Non-global logarithms  enter in a number of ways: in particular,
from the boundary at $\theta \sim R$, they affect the fat-jet mass,
and through it the distribution of the pruning radius.
They will affect both the $\sane$ and $\anomalous$ components
starting, in the small-$\zcut$ limit, from order $\as^3$.

The comparison between the analytic calculation and the Pythia shower is shown in Fig.~\ref{fig:analytic-vs-MC} in the middle panel. There one observes excellent agreement between the shapes of the analytical and MC distributions, indiacting once again a successful analytical description of pruning.

\subsection{MDT and mMDT mass distribution}\label{sec:mMDT}

The mass-drop tagger (MDT)~\cite{Butterworth:2008iy} is a declustering algorithm to be used with Cambridge/Aachen jets~\cite{Dokshitzer:1997in,Wobisch:1998wt}. 
In its original incarnation, the algorithm starts from a jet $j$, then undoes the last step of the clustering finding two subjets $j_1$ and $j_2$, with $m_{j_1}>m_{j_2}$
If there was a significant mass drop, $m_{j_1} < \mu m_{j}$, and
  the splitting is not too asymmetric, $y = \min(p_{tj_1}^2,
  p_{tj_2}^2) \Delta R_{j_1 j_2}^2/ m_j^2 > \ycut$, then the jet $j$ is tagged. Otherwise $j$ is redefined to be equal to
  $j_1$ and the algorithm iterates (unless $j$ consists of just a single
  particle, in which case the original jet is deemed untagged).
  
At $\order \as $ the mass-drop condition is always satisfied, so we only need to check for the $\ycut$ condition, which is essentially a cut on the energy sharing between the two prongs.  This situation is completely analogous to what we have encountered for pruning at LO and the resulting mass distribution has only a single logarithm.  However, starting from NLO the behaviour of MDT is far 
from straightforward. Complications arise because MDT recurses on the more massive branch, which in principle can be the softer of a given subjet pair. This was not what was intended in the original design, intended to tag hard substructure, and is to be considered a flaw.
We have in fact explicitly computed this {\it{wrong-branch}} contribution at NLO~\cite{DasFregMarSal,DasFregMarPow} and found that it generates a contribution to $\Sigma$ as singular as $\as^2 L^3$.  The ``wrong branch" contribution turns out to be numerically small but nevertheless calls for a modification.

The modified mass drop tagger (mMDT) is instead defined in such a way that it recurses on the subjet with the largest $m^2+p_t^2$. Not only does the mMDT eliminate the wrong-branch issue, but it also turns out to greatly facilitate the resummation of the tagged mass distribution. We find that the all-order mMDT mass distribution is simply given by the exponential of the one-loop result:
\begin{equation}
  \label{eq:modMDT-rho-ordering}
    \Sigma^\text{(mMDT)} (\rho)=
   \exp \left[ -D(\max(\ycut,\rho)) 
                  - S(\ycut,\rho) \Theta(\ycut - \rho) \right]\,.
  \end{equation}
The mass distribution above has  remarkable properties: it only contains single-logarithmic ($\as^nL^n$) contributions. All contributions from soft emissions have been successfully removed. It is to our knowledge the first time that a jet-mass type observable is found with this property. We will analyse the salient properties of mMDT in more detail in the next section.

The comparison between our analytic calculation and the Pythia shower is shown in Fig.~\ref{fig:analytic-vs-MC} in the bottom panel and yet again we note that our resummation perfectly captures the behaviour of the mMDT.

\section{Significant features of the modified mass drop tagger (mMDT)}\label{sec:mMDT-prop}
\begin{figure}\begin{center}
  \includegraphics[width=0.4\textwidth]{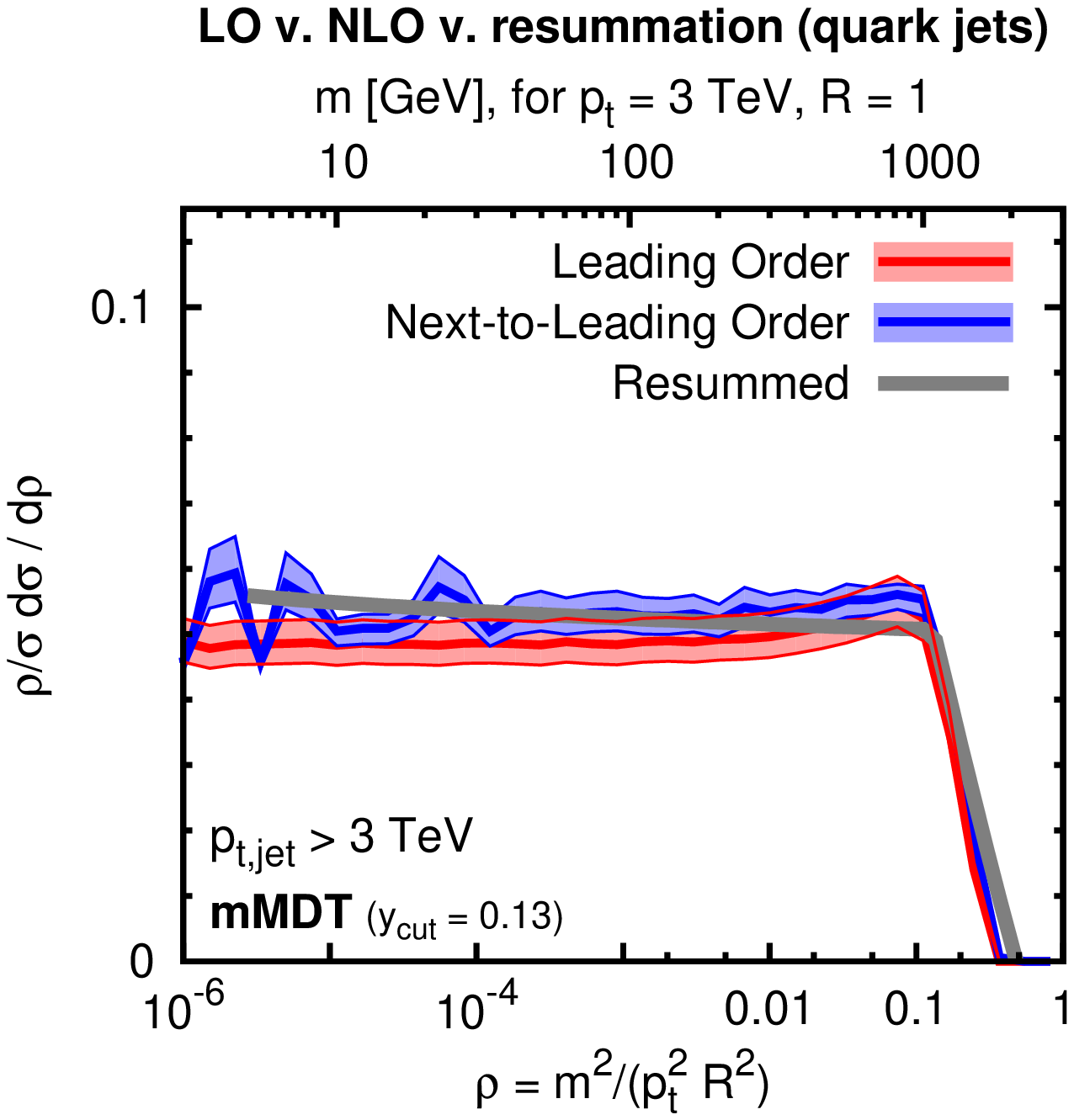}
  \includegraphics[width=0.4\textwidth]{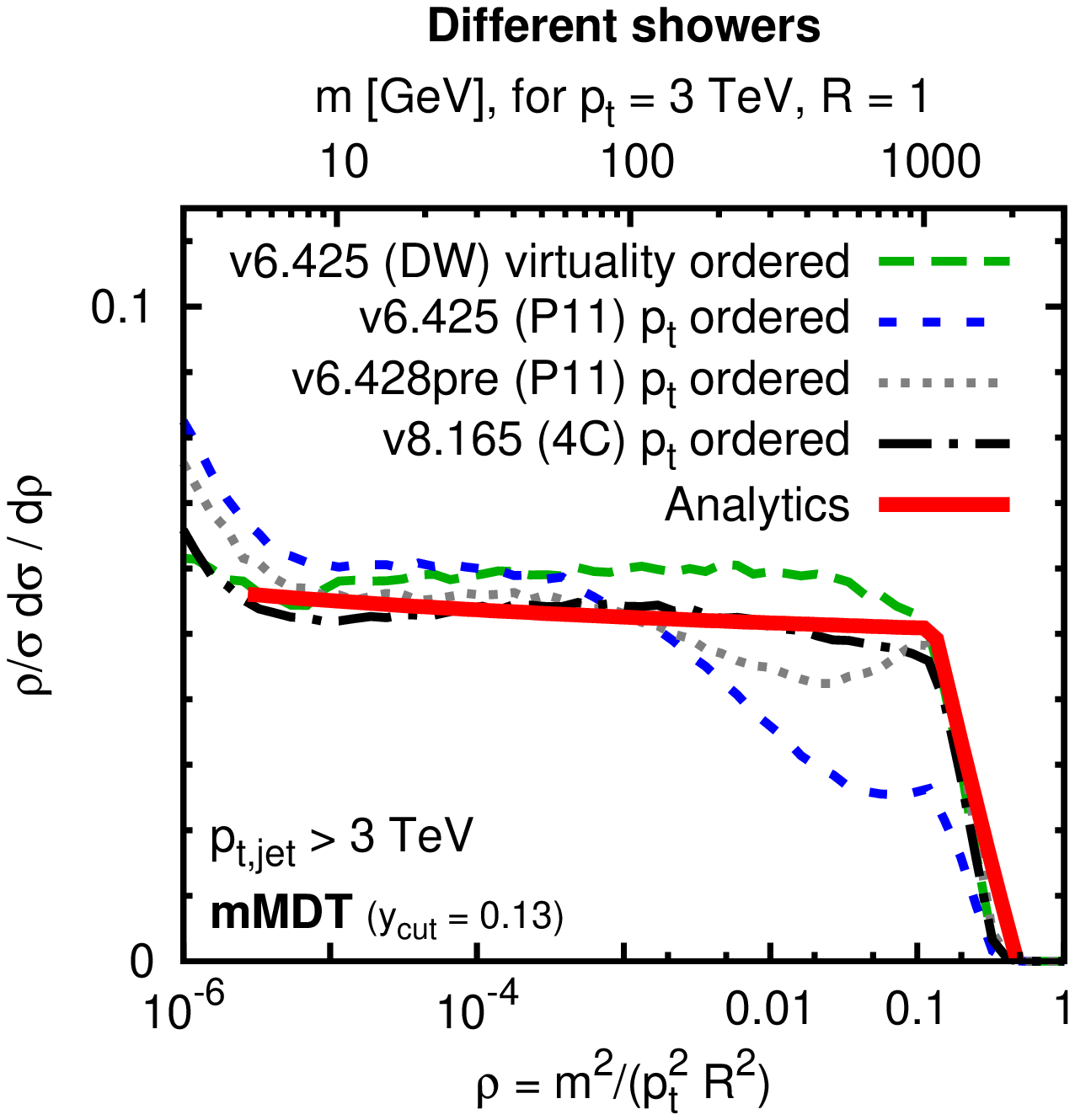}
    \caption{Comparison of the resummed result for mMDT to LO and NLO obtained with NLOJet++~\cite{Nagy:2003tz} (on the left) and to different Pythia parton showers (on the right).}  \label{fig:nlo-mc}
    \end{center}
\end{figure}

\begin{figure}\begin{center}
  \includegraphics[width=0.4\textwidth]{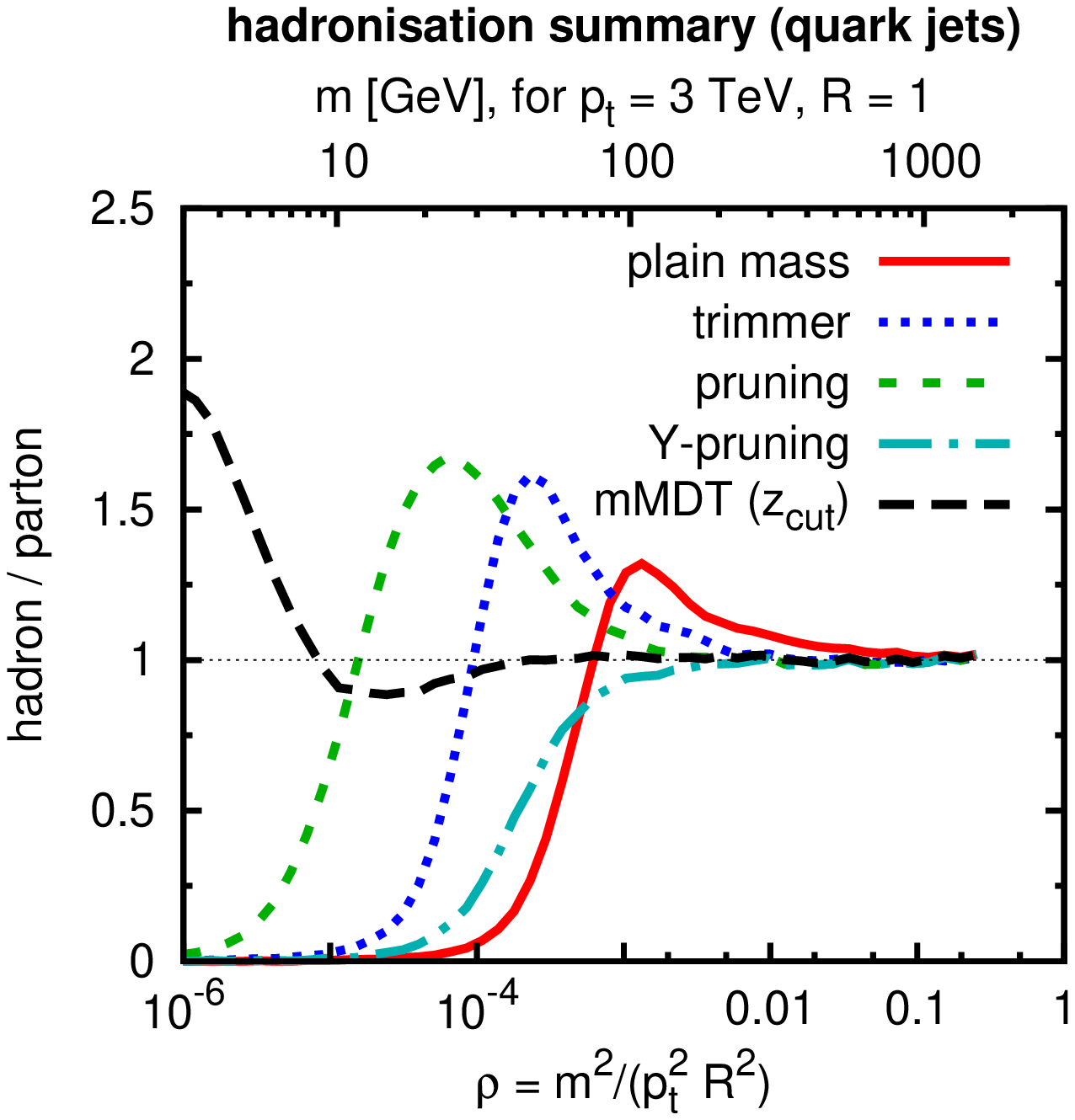}
  \includegraphics[width=0.4\textwidth]{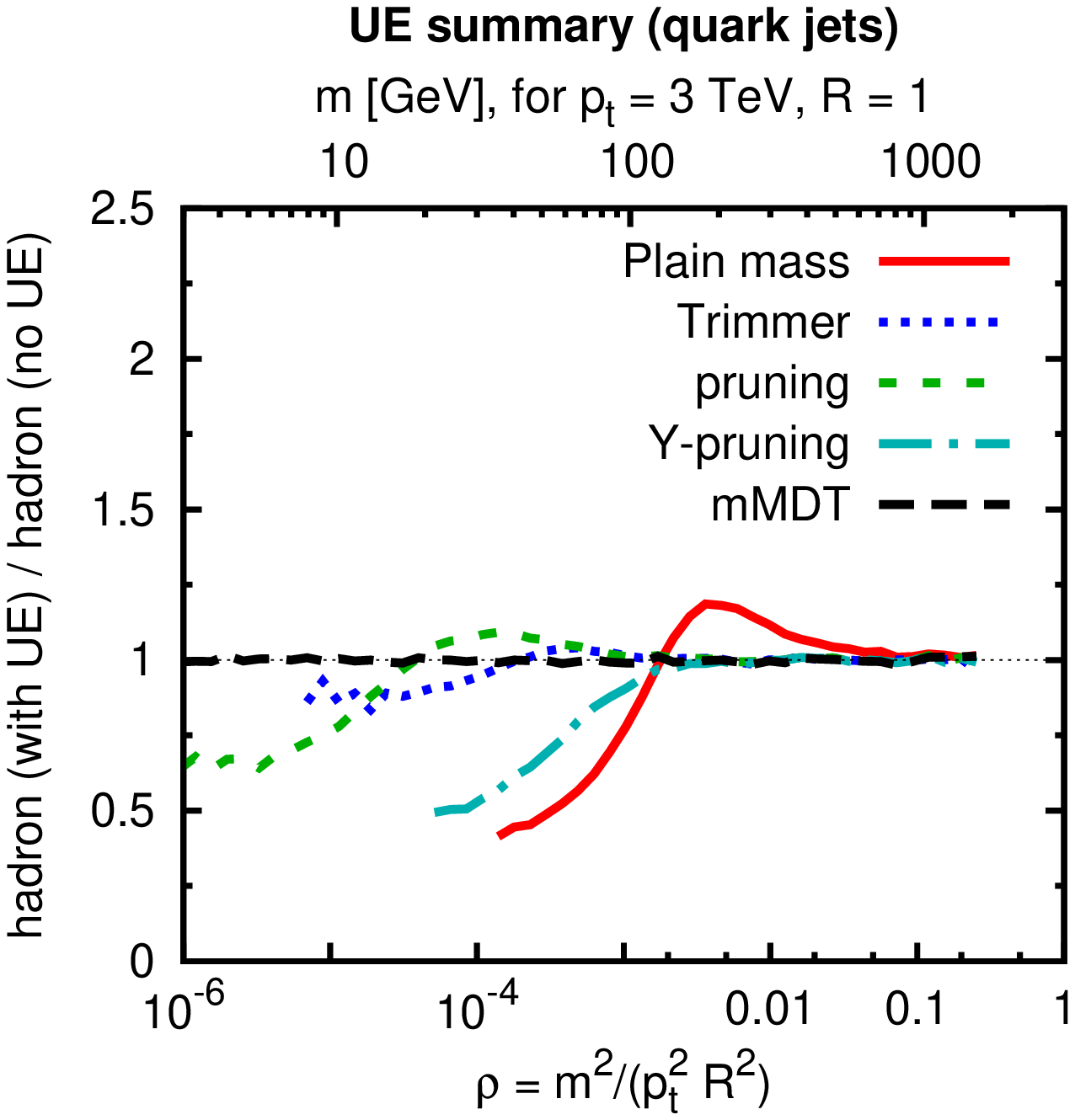}
  \end{center}
    \caption{Left: ratio of hadron-level (without UE) to parton level
    results for various groomers and taggers. 
    Right: ratio of hadron
    levels with and without UE.
    The details of the MC event generation are as for
    Fig.~\ref{fig:analytic-vs-MC}.}  \label{fig:NP-ratios}
\end{figure}

Given its remarkable properties it is worth summarising the main features of the mMDT.
\begin{itemize}
\item \emph{Background shapes}. mMDT mass distributions are free of Sudakov peaks and their shape is fairly insensitive to changes in $p_t$. Moreover, the value of $\ycut$ can be adjusted in order to obtain a flat distribution for the background, which is potentially advantageous for data driven background studies.
\footnote{Despite the name, the mass drop parameter $\mu$ does not significantly influence the shape of the distribution (if it is not taken too small). Also the mass-drop procedure is often used together with filtering~\cite{Butterworth:2008iy}, which only modifies the resummed result at the N$^{\nfilt}$LL level, which for the standard choice $\nfilt=3$ is highly subleading.}
\item \emph{Calculability}. 
\begin{enumerate}
\item \emph{Fixed order}
An interesting consequence of the presence of only single logarithms
relates to the validity of fixed-order perturbation theory, which is is expected to be valid down to $L \sim 1/\as$, rather than only down to  $ L \sim 1/\sqrt{\as}$. This is shown in Fig.~\ref{fig:nlo-mc}, on the left, where the resummed result is plotted together with fixed-order predictions.
\item \emph{Resummed level}.
 We have seen that mMDT completely removes contributions from soft emissions:
%i.e.\ one is left only with collinear divergences, but not
soft-collinear ones, or pure soft ones.
The absence of pure soft divergences has an important consequence, namely the absence of non-global logarithms. This makes the mMDT particularly interesting and it suggests that the mMDT should be given priority in calculations aiming for accuracy beyond single logarithms.
\item \emph{Non-perturbative corrections}.
So far we have concentrated on perturbative predictions. Clearly in the context of calculability we also need to take into account non perturbative effects. 
These include hadronisation, for which analytic estimates are perhaps possible, and underlying event contributions.
A Monte Carlo study of hadronisation effects and underlying event is summarised in Fig.~\ref{fig:NP-ratios}. One may expect groomers and taggers to have 
reduced sensitivity to non-perturbative physics. This is particularly striking for mMDT which for these values of $p_t$ has very small hadronisation 
corrections (not the case for pruning or trimming) and effectively no sensitivity to the underlying event.
\end{enumerate}
\end{itemize}
Therefore, we can conclude that mMDT not only provides a very useful tool for new physics searches (for which it was originally designed) but also appears to have special theoretical properties, which make it potentially of value for QCD measurements and studies including accurate $\alpha_s$ extraction. Additionally we can use it to probe and, perhaps, tune different Monte Carlo parton showers. A study in this direction is shown on the right-hand side of Fig.~\ref{fig:nlo-mc} where the resummed result is plotted together with different versions of the Pythia shower. 
The plot shows that
nearly all the Monte Carlo generators are in reasonable agreement with
each other and with our resummation.
The one exception is the $p_t$-ordered shower in Pythia~6.245, which
predicts a noticeably different shape for the distribution, both at
small and large masses. Following our calculations this discrepancy
was looked into by the Pythia authors, who after identifying an issue
in their shower, released a modified version, labelled v6.428pre,
which is in much better agreement with our analytics. 
This example illustrates the value of analytical understanding in
situations such as this where Monte Carlo results from various
generators differ noticeably.

%\section{Conclusions}

 \end{document}